\documentclass[journal,12pt,draftclsnofoot,onecolumn,english]{IEEEtran}

\usepackage{epsfig}
\usepackage{times}
\usepackage{float}
\usepackage{afterpage}
\usepackage{amsmath}
\usepackage{amstext,cite}
\usepackage{amssymb,bm}
\usepackage{latexsym}
\usepackage{color}
\usepackage{graphicx}
\usepackage{amsmath}
\usepackage{amsthm}
\usepackage{graphicx}
\usepackage[center]{caption}
\usepackage{pstricks}
\usepackage{caption}
\usepackage{subcaption}
\usepackage{booktabs}
\usepackage{multicol}
\usepackage{lipsum}
\usepackage[T1]{fontenc}
\usepackage{hyperref}
 \usepackage{aecompl}
 \usepackage{mathrsfs}

 \usepackage{arydshln}

\usepackage{soul}
\usepackage{cancel}

\usepackage{changes}
\definechangesauthor[color=red,name={Mingyue Ji}]{MJ}

\allowdisplaybreaks

\long\def\comment#1{}

\newcommand{\wv}{{\mathbf w}}

\newcommand{\Gc}{{\mathcal G}}

\newcommand{\Ic}{{\mathcal I}}

\newcommand{\Tc}{{\mathcal T}}

\newcommand{\gsf}{{\mathsf g}}

\newcommand{\qsf}{{\mathsf q}}

\newcommand{\Fsf}{{\mathsf F}}

\newcommand{\Ksf}{{\mathsf K}}
\newcommand{\Lsf}{{\mathsf L}}
\newcommand{\Msf}{{\mathsf M}}
\newcommand{\Nsf}{{\mathsf N}}

\newcommand{\Rsf}{{\mathsf R}}

\def\Fsf{{\mathsf F}}

\newtheorem{thm}{Theorem}

\newtheorem{lem}{Lemma}

\newtheorem{rem}{Remark}
\newtheorem{example}{Example}

\providecommand{\definitionname}{Definition}

\begin{document}

\title{Cache-aided General Linear Function Retrieval}  
\author{
Kai~Wan,~\IEEEmembership{Member,~IEEE,} 
Hua~Sun,~\IEEEmembership{Member,~IEEE,}
Mingyue~Ji,~\IEEEmembership{Member,~IEEE,}  
Daniela Tuninetti,~\IEEEmembership{Senior~Member,~IEEE,}
and~Giuseppe Caire,~\IEEEmembership{Fellow,~IEEE}
\thanks{
K.~Wan and G.~Caire are with the Electrical Engineering and Computer Science Department, Technische Universit\"at Berlin, 10587 Berlin, Germany (e-mail:  kai.wan@tu-berlin.de; caire@tu-berlin.de). The work of K.~Wan and G.~Caire was partially funded by the European Research Council under the ERC Advanced Grant N. 789190, CARENET.}
\thanks{
H.~Sun is with the Department of Electrical Engineering, University of North Texas, Denton, TX 76203 (email: hua.sun@unt.edu). The work of H.~Sun was supported in part by NSF Award 2007108.
}
\thanks{
M.~Ji is with the Electrical and Computer Engineering Department, University of Utah, Salt Lake City, UT 84112, USA (e-mail: mingyue.ji@utah.edu). The work of M.~Ji was supported in part by NSF Awards 1817154 and 1824558.}
\thanks{
D.~Tuninetti is with the Electrical and Computer Engineering Department, University of Illinois at Chicago, Chicago, IL 60607, USA (e-mail: danielat@uic.edu). The work of D.~Tuninetti was supported in part by NSF Award 1910309.}
}
\maketitle
\begin{abstract}
Coded Caching, proposed by Maddah-Ali and Niesen (MAN), has the potential to reduce network traffic by pre-storing content in the users' local memories when the network is underutilized and transmitting coded multicast messages that simultaneously benefit many users at once during peak-hour times. 
This paper considers the linear function retrieval version of the original coded caching setting, where users are interested in retrieving a number of linear combinations of the data points stored at the server, as opposed to a single file. This extends the scope of the Authors' past work that only considered the class of linear functions that operate element-wise over the files.
On observing that the existing cache-aided scalar linear function retrieval scheme does not work   in the proposed  
setting, this paper designs 
a novel coded caching scheme 
that outperforms uncoded caching schemes that either use unicast transmissions or let each user recover all files in the library. 
\end{abstract}
\begin{IEEEkeywords}
Coded caching; linear function retrieval; uncoded cache placement
\end{IEEEkeywords}

\section{Introduction}
\label{sec:intro}
Content caching is an efficient technique to handle the 
increase of requests for massive amounts of data and content over communication networks.   
By leveraging low-cost memory components at the user sides, caching reduces
peak-time traffic by prefetching contents closer to users
during off-peak time, thereby  reducing
the transmission delay or equivalently increasing the  bandwidth in   communication systems.
Traditional caching techniques aim at prefetching popular content by predicting the user demands, thus realizing a ``local caching gain'' (i.e., that scales with the amount of local memory)~\cite{uncodedcaching}. Maddah-Ali and Niesen (MAN) showed that it is possible to actually attain a ``global caching gain'' (i.e., that scales with the global amount of memory in the network) by using 
codes~\cite{dvbt2fundamental}.
The idea is that, if a single transmission can serve a number of  users simultaneously, the network load can be reduced by the same factor thus speeding-up communications significantly.
   
In the MAN setting, a server has a library of $\Nsf$ files and broadcasts to $\Ksf$ users through an error-free shared-link.
Each user has a cache of size of at most 
$\Msf$ files.
The MAN scheme consists of two phases: 
{\it placement phase}, where the server pushes content from the library to the local caches 
without knowledge of user future demands, and {\it delivery phase}, where each user requests one file and the server broadcasts coded packets such that each user can correctly recover its desired file. The objective is to minimize the {\it worst-case load} over all possible user demands, that is, the number of files that must 
be communicated so that any demands can be satisfied.
The MAN scheme is optimal under the constraint of uncoded cache placement (i.e., each user directly stores a collection of segments of the library files in its cache) when $\Nsf \geq \Ksf$ \cite{ontheoptimality,indexcodingcaching2020}. 
By removing the redundant transmissions in the MAN scheme when a file is requested multiple times, Yu, Maddah-Ali, and Avestimehr (YMA) derived a scheme that is optimal under the constraint of uncoded cache placement for $\Nsf < \Ksf$~\cite{exactrateuncoded}. In general, the YMA scheme is order optimal to within a factor of $2$~\cite{yufactor2TIT2018}, that is, coded placement can at best half the load of the YMA scheme.

 On the motivation that linear and multivariate polynomial queries naturally arise in modern engineering problems and  deep learning algorithms such as matrix-vector, matrix-matrix multiplications,  
 in~\cite{arxivfunctionretrieval} the Authors  posed the question of what is the optimal worst-case load when the cache-aided users are interested in retrieving a scalar linear function of the files rather than a single file. For the class of functions considered in~\cite{arxivfunctionretrieval}, which are restricted to operate element-wise on the file entries, it was surprisingly shown that the YMA load can be achieved, that is, there is no penalty in terms of load in retrieving scalar linear functions under the constraint of uncoded cache placement. 
It was noted in~\cite{arxivfunctionretrieval} that the proposed scalar linear function scheme can be  extended to all scenarios to which the original MAN
scheme has been extended, such as for example demand-private retrieval~\cite{wan2019privatecaching} and Device-to-Device networks~\cite{d2dcaching}. In addition, the scalar linear function scheme~\cite{arxivfunctionretrieval} can be used as a building block to provide demand-privacy and content-security against colluding users~\cite{colluding2020yan,security+privacy2020yan}.

In this paper, we move to a more general case of cache-aided linear function retrieval than in~\cite{arxivfunctionretrieval}, where users can request {\it general linear combinations} of  all symbols in the library, and not necessarily restricted to operate element-wise on the file entries.    For example, each user aims to compute some statistics of a bunch of data such as local weighted averages (which are general linear functions) of the data; these are   very common tasks in many applications depending on the data and on the weights.   

Besides the novel and realistic  problem formulation, our main contributions are as follows. 
We first introduce a baseline scheme that either lets each user recover all the  symbols in the library  or uses unicast transmissions to satisfy each user.  
The main challenge to implement a coded caching strategy in this problem is that  each symbol in a user's demand is a linear combination of all the  symbols in the library.
Inspired by the grouping coded caching strategy in~\cite{finiteanalysis}, which was used to reduce the sub-packetization level\footnote{The sub-packetization level is the smallest file length necessary to realize an achievable scheme.}, 
we propose a scheme that treats the demand of each user as a matrix-vector multiplication and uses the grouping strategy  to generate multicast messages after  possibly performing invertible linear matrix operations.  
The proposed scheme outperforms the baseline scheme in all parameter regimes.

\subsection*{Paper Organization}
The rest of this paper is organized as follows. 
Section~\ref{sec:system} formulates the  shared-link cache-aided general linear function retrieval problem.
Section~\ref{sec:main} provides the main result of this paper.
Section~\ref{sec:numerical} provides some numerical evaluations. 
Section~\ref{sec:conclusion} concludes the paper.
Some proofs may be found in Appendices.

\subsection*{Notation Convention}
Calligraphic symbols denote sets, 
bold symbols denote vectors and matrices,
and sans-serif symbols denote system parameters.
We use $|\cdot|$ to represent the cardinality of a set or the length of a vector;
$[a:b]:=\left\{ a,a+1,\ldots,b\right\}$ and $[n] := [1:n]$; 
$\oplus$ represents bit-wise XOR;  
$[a]^+:=\max\{a,0\}$; 
$\mathbb{F}_{\qsf}$ represents a  finite field with order $\qsf$;         
$\mathbf{A}^{\text{T}}$  and $\mathbf{A}^{-1}$ represent the transpose  and the inverse of matrix $\mathbf{A}$, respectively;
$\text{\rm rank}_q(\mathbb{A})$ represents the rank of matrix $\mathbb{A}$ on field   $\mathbb{F}_q$;
$\mathbf{I}_n$ represents the identity matrix with dimension $n \times n$;
$(\mathbf{A})_{m \times n}$ represents the dimension of matrix $\mathbf{A}$ is $m \times n$;  
  we let $\binom{x}{y}=0$ if $x<0$ or $y<0$ or $x<y$;

\section{System Model} 
\label{sec:system}
Different from~\cite{arxivfunctionretrieval}, here we consider the case where the users' desired linear functions are no longer scalar or operating element-wise across the files entries, thus we consider the whole library as a single file.

The  $(\Ksf,\Fsf,\Lsf,\qsf)$  shared-link cache-aided general linear function retrieval problem consists of a central server with access to a library of $\Fsf$ 
independent and identically distributed (i.i.d.)
  symbols over a finite filed $\mathbb{F}_{\qsf}$,  denoted by $\mathbf{w}=(w_1,\ldots,w_\Fsf)^T \in (\mathbb{F}_{\qsf})^{\Fsf}.$  
We often treat $\mathbf{w}$ as a column vector, which should be clear from the context.  
The server is connected to $\Ksf$ cache-aided users through an error-free shared-link. 
 The system has two phases.
\begin{itemize}

\item
In the {\bf placement phase}, the server pushes up to $\Msf$  
symbols into the local cache of each user, where $\Msf \in [0:\Fsf]$, 
without knowing what the users will   demand later.
The cached content of user $k\in[\Ksf]$ is denoted by 
\begin{align}
Z_{k}=\phi_{k}(\mathbf{w}),  
\label{eq:cK}
\end{align}  
where $\phi_k$ is the placement function for user $k$ defined as
\begin{align}
\phi_k &: (\mathbb{F}_{\qsf})^{\Fsf} \to (\mathbb{F}_{\qsf})^{\Msf}, \quad k\in[\Ksf].
\label{eq: placement functions def}
\end{align} 
$\Msf$ is referred to as the cache  (or memory) size.
If each user directly copies $\Msf$ symbols from the library into its cache, the cache placement is said to be {\it uncoded}.

\item
In the {\bf delivery phase}, each user wants to retrieve $\Lsf$ 
linear combinations of all the symbols in the library, where $\Lsf \in [1:\Fsf]$. 
The demand of user $k\in [\Ksf]$  is represented by the matrix $\mathbf{D}_k \in (\mathbb{F}_{\qsf})^{\Lsf \times \Fsf}$, meaning    user $k$ aims to retrieve   
\begin{align}
\mathbf{y}_k = \mathbf{D}_k \ \mathbf{w} \in (\mathbb{F}_{\qsf})^{\Lsf},  
\label{eq:demand of k}
\end{align}
Let the collection of all demand matrices be $\mathbf{D} := [\mathbf{D}_1;\ldots;\mathbf{D}_{\Ksf}] \in (\mathbb{F}_{\qsf})^{\Ksf\Lsf \times \Fsf}$.
We assume that the server and all users know $\mathbf{D}$ which is communicated on a separate channel, thus not impacting the downlink load next -- see also Remark~\ref{rem:up}.\footnote{\label{note:different from matrix}
Notice that differently from the cache-aided matrix multiplication problem in~\cite{cachematrixmultiplication2020}, where the matrix on the each side of the  desired multiplication is one of the library files, in this paper each user $k\in[\Ksf]$ desires $\mathbf{D}_k  \mathbf{w}$ where $\mathbf{D}_k$ is known by all the users in the delivery phase and $ \mathbf{w}$ represents the vector of all symbols in the library.}

According to all the users' demand matrix $\mathbf{D}$, the server broadcasts the message 
\begin{align}
 X = \psi(\mathbf{D}, \mathbf{w}), 
\label{eq:psi}
\end{align}
where $\psi$ is the encoding function 
\begin{align}
\psi &: (\mathbb{F}_{\qsf})^{\Ksf\Lsf \times \Fsf} \times (\mathbb{F}_{\qsf})^{\Fsf} \to (\mathbb{F}_{\qsf})^{\Rsf},
\label{eq: encoding function def}
\end{align}
for some $\Rsf \in [0:\Fsf]$. 
$\Rsf$ is referred to as the load.

\end{itemize}


{\it Achievability:}
For the $(\Ksf,\Fsf,\Lsf,\qsf)$ shared-link cache-aided general linear function retrieval problem, 
we say that   the pair $(\Msf,\Rsf)$ is achievable if for any possible demand $\mathbf{D}$  there exist placement functions in~\eqref{eq: placement functions def} and
a delivery function in~\eqref{eq: encoding function def} such that 
\begin{align}
H(\mathbf{D}_k \mathbf{w} | \mathbf{D}, Z_k, X) = 0, \quad \forall k\in[\Ksf].
\label{eq: decoding functions def}
\end{align}

{\it Optimal memory-load tradeoff:}
For the $(\Ksf,\Fsf,\Lsf,\qsf)$ shared-link cache-aided general linear function retrieval problem, 
the objective is to determine the minimum worst-case downlink load  (or   load for simplicity)
defined as  
\begin{align}
  \Rsf^{\star}(\Msf) =  
  \min_{ \phi_1, \ldots,\phi_\Ksf, \psi }  
  \{\Rsf : \text{$(\Msf,\Rsf)$ is achievable} \}.
\label{eq:def of worst case load ambitious}
\end{align}

{\it Optimal memory-load tradeoff in the limit for large file size:}
Since solving the problem in~\eqref{eq:def of worst case load ambitious} for any given $(\Ksf,\Fsf,\Lsf,\qsf)$ is challenging, in the following we shall consider the regime where the file size $\Fsf$ is as large as desired and we thus let the system parameters scale with the file length as follows
\begin{align}
\Msf &:= \mu \Fsf, \ \mu\in[0,1], \\
\Lsf &:= \lambda \Fsf, \ \lambda\in[0,1], \\
\Rsf &:= \rho \Fsf, \ \rho\in[0,1].
\end{align}
For fixed $(\Ksf,\lambda)$ we aim to characterize the minimum worst-case   normalized downlink load   (or normalized load for simplicity)
\begin{align}
\rho^\star(\mu) = \min_{\phi_1, \ldots,\phi_\Ksf, \psi } 
\{ \rho :   \text{$(\Msf,\Rsf)=(\mu \Fsf,\rho \Fsf)$ is achievable for {\it some} $(\Fsf,\qsf)$} \}.
\label{eq:def of worst case load}
\end{align}

\begin{rem}[Relationship to~\cite{arxivfunctionretrieval}]\label{rem:cover scalar function retrieval}
The cache-aided scalar linear function retrieval problem  in~\cite{arxivfunctionretrieval}  is a special case of the formulation here. 
More precisely, let $\Fsf = \Nsf\Lsf$ (i.e.,  $\frac{1}{\Nsf}=\lambda$),    
  where $\Nsf$ indicates the number of files and $\lambda \Fsf $ is the file length. 
The demand of user $k\in[\Ksf]$ is represented by the vector $\mathbf{y}_k=(y_{k,1}, y_{k,2}, \ldots, y_{k,\Nsf}) \in (\mathbb{F}_{\qsf})^{\Nsf}$ by which we mean that the user is requesting 
\begin{align}
\mathbf{D}_k = 
\begin{bmatrix}
y_{k,1}\mathbf{I}_{\Lsf}, \ y_{k,2}\mathbf{I}_{\Lsf}, \ldots, y_{k,\Nsf}\mathbf{I}_{\Lsf}
\end{bmatrix} \in (\mathbb{F}_{\qsf})^{\Lsf \times \Nsf\Lsf}, 
\label{eq:qazxsw}
\end{align}
where $\mathbf{I}_n$ is the identity matrix with dimension $n \times n$.
In the restricted setting where the demands are as in~\eqref{eq:qazxsw} the optimal  load under the constraint of uncoded cache placement is the lower convex envelop of the points
\begin{subequations}
\begin{align}
 \left(\frac{\Msf}{\Lsf}, \frac{\Rsf_{\text{scalar}}}{\Lsf}\right)
 &= \left(\frac{\Nsf \ t}{\Ksf}, \ 
  \frac{\binom{\Ksf}{t+1} - \binom{\Ksf- \min\{\Ksf,\Nsf \} }{t+1} }{\binom{\Ksf}{t}} 
  \right), t\in[0:\Ksf], 
\label{eq:YMA} \\
     \Longleftrightarrow \ (\mu, \rho_{\text{scalar}})  &= \left(\frac{\ t}{\Ksf}, \ 
  \lambda \frac{\binom{\Ksf}{t+1} - \binom{\Ksf- \min\{\Ksf,\Nsf \} }{t+1} }{  \binom{\Ksf}{t}} 
  \right), t\in[0:\Ksf], \label{eq:YMA rho} 
\end{align}
\end{subequations}
where for a given value of $t$ in~\eqref{eq:YMA} the subpacketization level $\Lsf$ must be an integer multiple of $\binom{\Ksf}{t}$. 
%
\hfill $\square$ 
\end{rem}

\begin{rem}[A minrank solution]\label{rem:minrank}
For the $(\Ksf,\Fsf,\Lsf,\qsf)$ shared-link cache-aided general linear function retrieval problem, 
the best linear scheme, inspired by~\cite{indexcodingwithsi,codedsideinfor}, is a follows.
Linear placement: user $k\in[\Ksf]$ caches $Z_k=\mathbf{P}_k \mathbf{w} \in (\mathbb{F}_\qsf)^{\Msf}$ for some $\mathbf{P}_k \in (\mathbb{F}_\qsf)^{ \Msf \times \Fsf}$.
Linear delivery: 
the server sends, in the worst case, a number of symbols given by  
\begin{align}
\Rsf_{\text{minrank}} = 
\min_{\mathbf{P}_1,\ldots,\mathbf{P}_K} 
\max_{\mathbf{D}_1,\ldots,\mathbf{D}_K} 
\min_{\mathbf{T}_1,\ldots,\mathbf{T}_K} 
{\rm rank} 
\begin{bmatrix}
\mathbf{D}_1 + \mathbf{T}_1 \mathbf{P}_1\\ 
\mathbf{D}_2 + \mathbf{T}_2 \mathbf{P}_2\\ 
\vdots \\
\mathbf{D}_\Ksf + \mathbf{T}_\Ksf \mathbf{P}_\Ksf\\ 
\end{bmatrix},
\label{eq:minrank}
\end{align}
where $\mathbf{T}_k  \in (\mathbb{F}_\qsf)^{ \Lsf \times \Msf}, \ k\in[\Ksf]$.
Solving the minrank problem in~\eqref{eq:minrank} is hard~\cite{indexcodingwithsi,codedsideinfor}, thus in the following we shall design a scheme with lower complexity.
\hfill $\square$ 
\end{rem}

\begin{rem}[A baseline scheme]\label{rem:baseline}
For the $(\Ksf,\Fsf,\Lsf,\qsf)$ shared-link cache-aided general linear function retrieval problem, the load 
\begin{subequations}
\begin{align}
 \Rsf_{\text{baseline}} &=  \min\left\{  \Ksf  \Lsf  ,  \Fsf -\Msf \right\}
 \\ \Longleftrightarrow \ \rho_{\text{baseline}} &=  \min\left\{  \Ksf  \lambda  , 1-\mu \right\},
\end{align} 
 \label{eq:first baseline load} 
\end{subequations}
can be achieved by an uncoded caching strategy  as follows.
\begin{itemize}

 \item In order to achieve the load $ \Ksf   \Lsf$, 
  we transmit one by one the elements of $\mathbf{y}_k, k\in[\Ksf],$ in~\eqref{eq:demand of k}.
  The main limitation of this unicast transmission scheme 
  is the lack of multicast gain.
  
 \item In order to achieve $ \Fsf  -\Msf$ 
 we let each user recover all the symbols in the library. In the placement phase, each user caches the first $\Msf$ 
 symbols in the library. In the delivery phase, the server transmits all the remaining  $ \Fsf -\Msf$ 
 symbols. 
 The main limitation of this scheme is that, if $\Lsf <  \Fsf -\Msf$, 
 the users do not need to recover all the symbols in the library in order to retrieve their desired function. 
 
\end{itemize}
The main contribution of this paper is to find schemes that, despite the lack of structure on the demand matrices in general, achieve a smaller load than~\eqref{eq:first baseline load}.
\hfill $\square$ 
\end{rem}

\begin{rem}[Uplink and downlink loads]\label{rem:up}
Besides downlink load, uplink load 
 is also considered in the distributed  matrix-matrix multiplication problem~\cite{jia2019matrixCSA,chang2019privatesecure,updowlink2019kakar}. 
In this work, the communication cost of uploading the demand matrix to the server is not a focus, i..e,   we assume that each user communicates the whole demand matrix to the server and all other users on a separate channel that is not the bottleneck in the system. 
%
This assumption can be also justified as follows. 
Let $\mathscr{D}(k)$ denotes the set of possible demand matrices of user $k\in[\Ksf]$, referred to as {\it demand range}, that is, user $k$ chooses one   matrix in $ \mathscr{D}(k)$ as its demand. 
We assume that $\mathscr{D}(k)$ is known by the server and all users. 
The communication cost to let the server and the other users know the realization of the demand matrix is negligible compared to the number of transmissions from the server if $\sum_{k\in[\Ksf]} \log_{\qsf} (|\mathscr{D}(k)|) \ll \Fsf$. 
\hfill $\square$ 
 \end{rem}

\section{Main Result}
\label{sec:main}


Based on Remark~\ref{rem:baseline}, the main challenge is to design a coded caching strategy that
(i) lets each user directly recover the desired linear combinations, instead of recovering all the library symbols, and
(ii) attains coded caching gain, as opposed to serving the users one-by-one with unicast transmissions.
The main contribution of this paper is 
the following theorem, which is proved in Appendix~\ref{app:achieve}.

\begin{thm}
\label{thm:main achievable scheme} 
For the $(\Ksf,\lambda)$ shared-link cache-aided general linear function retrieval problem, we have: 
\begin{itemize}
 \item if   $\mu =\alpha \frac{g-1}{g} +(1-\alpha) \frac{g}{g+1} $ where $g \in [\Ksf-1]$ and $\alpha \in [0,1]$,
the following  normalized load is achievable
 \begin{subequations}
\begin{align}
 \rho_{\text{ach}} &:= 
\begin{cases}
\left\lceil \frac{\Ksf}{g}\right\rceil  \lambda , & \ \text{ if } \frac{\alpha  }{\gsf} \geq \left\lceil \frac{\Ksf}{g}\right\rceil\lambda  \\ 
& \\
\min \left\{  \rho_1 , \rho_2 \right\} 
   &  \  \text{ if }  \frac{\alpha  }{\gsf} \leq \left\lceil \frac{\Ksf}{g}\right\rceil \lambda  
\end{cases},
\label{eq:first reg proposed ache}
\\
  \rho_1 &:= \frac{\alpha  }{g}  + \min\left\{ \left\lceil \frac{\Ksf}{g+1}\right\rceil  \lambda  , \frac{(1-\alpha) }{g+1} \right\} ,  \label{eq:r1} \\
  \rho_2 &:=\left\lceil \frac{\Ksf}{g}\right\rceil   \min\left\{\frac{\alpha  }{g} ,\lambda \right\}+  \min\left\{ \left\lceil \frac{\Ksf}{g+1}\right\rceil \left[\lambda-  \frac{\alpha  }{g}\right]^+ , \frac{(1-\alpha) }{g+1} \right\} ; \label{eq:r2}
 \end{align}
\label{eq:mainthm}
\end{subequations}
 \item  if   $\mu =\alpha \frac{\Ksf-1}{\Ksf} +(1-\alpha) $ where $\alpha \in [0,1]$,
  the following  normalized load is achievable
\begin{align}
\rho_{\text{ach}} = \rho_3 = \min\left\{ \frac{\alpha  }{\Ksf} ,\lambda \right\} . \label{eq:sec reg proposed ache dt3}
 \end{align}
 \end{itemize}
 \hfill $\square$ 
\end{thm}

Next, we provide  the intuition behind 
the proposed scheme in Theorem~\ref{thm:main achievable scheme}, which is based on three ingredients:
\begin{enumerate}

\item
We start by the achievable scheme for~\eqref{eq:sec reg proposed ache dt3} with $\alpha=1$. We aim to design the cache placement 
such that each user  caches  a fraction $\frac{\Ksf-1}{\Ksf} $ of the file and the uncached part of file by  this user is known by the remaining $\Ksf-1$ users. 
With this cache placement, the delivery consists of a single multicast message with multicasting gain $\Ksf$.  More precisely, the construction of the proposed scheme is as follows.

Assume $\Ksf$ divides $\Fsf$. 
We use here a Matlab-like notation for submatrices.
The library is partitioned into $\Ksf$ equal length subfiles as follows
\begin{align}
  \Ic_k   &:=   \left[(k-1)\frac{\Fsf}{\Ksf}+1: k \frac{\Fsf}{\Ksf}\right], k\in[\Ksf], 
\\ \mathbf{w}_{k} &:=  \mathbf{w}(\Ic_k) \in (\mathbb{F}_{\qsf})^{\frac{\Fsf}{\Ksf}}, k\in[\Ksf], 
\\ \mathbf{w} &= (\mathbf{w}_{1}, \ldots, \mathbf{w}_{\Ksf});
\end{align}
user $k\in[K]$ caches $Z_k = (\mathbf{w}_{j} : j\in[\Ksf]\setminus\{k\})$;
the server delivers the multicast message 
\begin{align}
X = 
\begin{cases}
\sum_{k\in[\Ksf]} \mathbf{D}_{k; \ :,\Ic_k }  \mathbf{w}_{k} \in (\mathbb{F}_{\qsf})^{\Lsf},  & \text{if } \frac{\Fsf}{\Ksf} > \Lsf \\
\sum_{k\in[\Ksf]} \mathbf{w}_{k} \in (\mathbb{F}_{\qsf})^{\frac{\Fsf}{\Ksf}} , & \text{if } \frac{\Fsf}{\Ksf} \leq \Lsf \\
\end{cases},
\end{align}
where $\mathbf{D}_{k; \ :,\Ic_k }$ represents the sub-matrix of $\mathbf{D}_{k}$ including the columns with indices in $\Ic_k$. In $X$, each user $k\in [\Ksf]$ knows
all but the requested vector
\begin{align*}
\begin{cases}
\mathbf{D}_{k; \ :,\Ic_k }  \mathbf{w}_{k} ,  &  \text{if }  \frac{\Fsf}{\Ksf} > \Lsf; \\
\mathbf{w}_{k}, & \text{if } \frac{\Fsf}{\Ksf} \leq \Lsf, \\
\end{cases},
\end{align*}
 such that user $k$ can recover either of them.  
Thus an achieved normalized memory-load tradeoff 
is
\begin{align}
(\mu,\rho) 
= \left(
1-\frac{1}{\Ksf}, \
\min\left\{\frac{1}{\Ksf} , \lambda \right\}
\right).
\label{eq:secondtolastcorner K-1} 
\end{align}

\item
 We then introduce the    achievable scheme for~\eqref{eq:first reg proposed ache} with $\alpha \in \{0,1\}$.  Assume now the $\Ksf$ users are portioned into $g$ groups of $\left\lceil \frac{\Ksf}{g}\right\rceil$ users each, where $g\in [\Ksf-1]$. 
Let the users in the same group share the same cache content and recover all the linear combinations demanded by the users in the group.
Then the normalized memory-load tradeoff is as in~\eqref{eq:secondtolastcorner K-1} 
but with $\Ksf$ replaced by with $g$ and $\Lsf$ replaced by $\left\lceil \frac{\Ksf}{g}\right\rceil\Lsf$. 
Therefore, we get that the following normalized memory-load points are achievable
\begin{align}
(\mu,\rho) 
 = \left(
1-\frac{1}{g}, \
\min\left\{\frac{1}{g} , \lambda \left\lceil \frac{\Ksf}{g}\right\rceil  \right)
\right), \ g\in[\Ksf].
\label{eq:secondtolastcornerG}
\end{align}

\item
The rest of the proof of Theorem~\ref{thm:main achievable scheme} consists of a method to `interpolate' among the points in~\eqref{eq:secondtolastcornerG}, as explained in   Appendix~\ref{app:achieve}. 
Unlike cache-aided scalar linear function retrieval  in~\cite{arxivfunctionretrieval}, the difficulty in the considered problem is that 
connecting two normalized memory-load points by a  line segment is generally impossible.
The main  reason is that  if we
partition $\wv$ as $\wv=[\wv^{\prime}; \wv^{\prime\prime}]$ and use a different cache placement strategy on each part, 
each demanded function $\mathbf{D}_k \wv$ is in the form
\begin{align}
\mathbf{D}_k \wv = \mathbf{D}^{\prime}_{k} \wv^{\prime} +  \mathbf{D}^{\prime\prime}_{k} \wv^{\prime\prime};
\end{align}
thus it cannot be divided into two separate parts, where the first part only contains the linear combinations of $\wv^{\prime}$ and the second part only contains the linear combinations of $\wv^{\prime\prime}$.
An example to highlight this limitation and our approach to overcome it is provided at the end of this section.
\end{enumerate}

\begin{rem}[Comparison to the baseline scheme]
\label{rem:better than baseline}
We  show here that  the proposed scheme in Theorem~\ref{thm:main achievable scheme}  outperforms the baseline scheme in~\eqref{rem:baseline}.  
\begin{itemize}

\item
Case  $\mu =\alpha \frac{g-1}{g} +(1-\alpha) \frac{g}{g+1} $  where $g\in [\Ksf]$ and $\alpha \in [0,1]$:
From~\eqref{eq:first reg proposed ache}  and~\eqref{eq:r2}, it can be seen that 
\begin{align}
\rho_{\text{ach}} \leq   \left\lceil \frac{\Ksf}{g}\right\rceil  \lambda \leq \Ksf \lambda. \label{eq:outperform case11}
\end{align} 
 From~\eqref{eq:first reg proposed ache} and~\eqref{eq:r1}, it can be seen that 
\begin{align}
\rho_{\text{ach}} \leq  \frac{\alpha  }{g}  + \frac{ 1-\alpha} {g+1} = 1-\mu. \label{eq:outperform case12}
\end{align} 
Hence, from~\eqref{eq:outperform case11} and~\eqref{eq:outperform case12}, we can prove   $\rho_{\text{ach}}  \leq \rho_{\text{baseline}} $ in this case.

\item
Case $\mu =\alpha \frac{\Ksf-1}{\Ksf} +(1-\alpha)  $ where $\alpha \in [0,1]$:
Since in this case $\frac{\alpha }{\Ksf}=1-\mu$, from~\eqref{eq:sec reg proposed ache dt3} we can prove $\rho_{\text{ach}}  \leq \rho_{\text{baseline}} $ in this case.

\end{itemize}
\hfill $\square$ 
\end{rem}

\begin{rem}[Connection to Remark~\ref{rem:cover scalar function retrieval}]
For the proposed scheme achieving~\eqref{eq:secondtolastcorner K-1}, the cache placement is the same as  the cache-aided scalar linear function retrieval scheme  in Remark~\ref{rem:cover scalar function retrieval} with $t=\Ksf-1$.

Notice that,  for the considered cache-aided general linear function retrieval problem where    $\mu=\frac{t}{\Ksf}$ and $t\in[\Ksf]$, we could use the cache-aided scalar linear function retrieval scheme  in Remark~\ref{rem:cover scalar function retrieval} to deliver $\binom{\Ksf}{t+1}$ pieces of the requested vectors.  
The scheme would achieve
\begin{align}
(\mu,\rho) 
  &= \left(
\frac{t}{\Ksf}, \ 
  \lambda \binom{\Ksf}{t+1} 
\right), t\in[\Ksf],
\label{eq:secondtolastcorner t}
\end{align}
which reduces to~\eqref{eq:secondtolastcorner K-1} for $t=\Ksf-1$.
By the grouping argument we would achieve
\begin{align}
(\mu,\rho^{\prime}) 
  &= \left(
\frac{t}{g}, \ 
  \lambda \left\lceil \frac{\Ksf}{g}\right\rceil \binom{g}{t+1} 
\right), t\in[g], g\in[\Ksf].
\label{eq:secondtolastcorner t g}
\end{align}
Let then fix one $g \in [\Ksf]$ and one $t \in [g-2]$, and  
analyse the achieved normalized load in~\eqref{eq:secondtolastcorner t g}.   We will show that 
\begin{align}
\rho^{\prime}=  \lambda \left\lceil \frac{\Ksf}{g}\right\rceil \binom{g}{t+1}    \geq \rho_{\text{baseline}} .\label{eq:worse than baseline}
\end{align}
as follows.
It can be seen that 
\begin{subequations}
\begin{align}
& \lambda \left\lceil \frac{\Ksf}{g}\right\rceil \binom{g}{t+1} \geq \Ksf \lambda  \frac{ \binom{g}{t+1}}{g}  \\
& \geq \Ksf \lambda  
\label{eq:binom g choose t1}
\\& \geq 
\rho_{\text{baseline}},
\end{align} 
\end{subequations}
where~\eqref{eq:binom g choose t1} follows since $t \in [g-2]$ and thus  $\binom{g}{t+1} \geq g$. 
This shows that, with the exception for the normalized memory-load points with $t=g-1$, the scheme in~\eqref{eq:secondtolastcorner t g} is inferior to  the baseline scheme in~\eqref{eq:first baseline load}, and will thus not be pursued in the rest of the paper. 
\hfill $\square$ 
\end{rem}

We finish this section with an example to illustrate the main ideas of the proposed scheme.

\begin{example}\rm
\label{ex:example alpha>0}
We consider a system 
with $\Ksf=6$ users, cache fraction $\mu =\frac{47}{72}$, and demand fraction $\lambda=\frac{1}{12}$.
It can be seen that 
\begin{align}
 \mu=\frac{47}{72} = \left. \alpha \frac{g-1}{g} +(1-\alpha) \frac{g}{g+1}  \right|_{\alpha= \frac{1}{12}, \ g=2}.
\end{align}

{\it  Placement Phase.}  
It can be seen that the memory size is between $ \mu_1= \frac{g-1}{g}  = \frac{1}{2}$  and $\mu_2=\frac{g}{g+1}= \frac{2}{3}$. 
We partition $\mathbf{w}$ into two parts as $\mathbf{w}=[\mathbf{w}^1 ; \mathbf{w}^2]$ where 
$\mathbf{w}^1 \in (\mathbb{F}_\qsf)^{\Fsf/12}$ 
and $\mathbf{w}^2\in (\mathbb{F}_\qsf)^{11\Fsf/12}$. 
Furthermore, 
\begin{itemize}
\item
$\mathbf{w}^1$ is  partitioned  into two equal-length subfiles, $\mathbf{w}^1=[\mathbf{w}^1_{\{1\}};\mathbf{w}^1_{\{2\}} ]$, each of which has  $\frac{\Fsf}{24}$ symbols. We divide   the $6$ users into $2$ groups where $\Gc^1_1=\{1,3,5\}$ and $\Gc^1_2=\{2,4,6\}$.
 We let users in $\Gc^1_1$ cache $\mathbf{w}^1_{\{1\}}$ and let users in $\Gc^1_2$ cache  $\mathbf{w}^1_{\{2\}}$.
\item
$\mathbf{w}^2 $ is partitioned  into three equal-length subfiles, $\mathbf{w}^2 =[ \mathbf{w}^2_{ \{1,2\}};\mathbf{w}^2_{ \{1,3\}};\mathbf{w}^2_{ \{2,3\}}]$, each of which has $\frac{11\Fsf}{36}$ symbols.  We divide the $6$ users into $3$ groups, where $\Gc^2_1=\{1,4\}$, $\Gc^2_2=\{2,5\}$, and $\Gc^2_3=\{3,6\}$.
 We let users in $\Gc^2_1$ cache   $\mathbf{w}^2_{ \{1,2\}}$ and $ \mathbf{w}^2_{ \{1,3\}}$, let users in $\Gc^2_2$ cache  $\mathbf{w}^2_{ \{1,2\}}  ,\mathbf{w}^2_{ \{2,3\}}$, and let users in  $\Gc^2_3$ cache  $\mathbf{w}^2_{ \{1,3\}}$ and $\mathbf{w}^2_{ \{2,3\}}$.
\end{itemize}
Each user caches $\frac{\Fsf}{24} + \frac{2 \times 11\Fsf }{36}  = \frac{47\Fsf}{72}$ symbols, thus satisfying the memory size constraint.

{\it  Delivery Phase.}  
With some permutation on the rows of  $\mathbf{w}$, the demand of user $1$ can be expressed as
 \begin{subequations}
 \begin{align}
 \mathbf{D}_1 \ \mathbf{w} 
=&\mathbf{D}_{1,\{1\}} \ \mathbf{w}^1_{\{1\}}  + \mathbf{D}_{1,\{1,2\}} \ \mathbf{w}^2_{\{1,2\}} +\mathbf{D}_{1,\{1,3\}}  \ \mathbf{w}^2_{\{1,3\}} + \mathbf{D}_{1,\{2\}} \ \mathbf{w}^1_{\{2\}}  + \mathbf{D}_{1,\{2,3\}} \ \mathbf{w}^2_{\{2,3\}} .\label{eq:example demand of user 1}
\end{align}
\end{subequations}
User $1$ can recover $\mathbf{D}_{1,\{1\}}   \mathbf{w}^1_{\{1\}}  + \mathbf{D}_{1,\{1,2\}}  \mathbf{w}^2_{\{1,2\}} +\mathbf{D}_{1,\{1,3\}}   \mathbf{w}^2_{\{1,3\}}$ from its cache, and similarly for the other users. Thus in the delivery phase, the users need to recover 
\begin{align}
\mathbf{B}_1:= \mathbf{D}_{1,\{2\}} \ \mathbf{w}^1_{\{2\}}  + \mathbf{D}_{1,\{2,3\}} \ \mathbf{w}^2_{\{2,3\}},\label{eq:user 1 needs to recover} \\
\mathbf{B}_2:=\mathbf{D}_{2,\{1\}} \ \mathbf{w}^1_{\{1\}}  + \mathbf{D}_{2,\{1,3\}} \ \mathbf{w}^2_{\{1,3\}},\label{eq:user 2 needs to recover} \\
\mathbf{B}_3:=\mathbf{D}_{3,\{2\}} \ \mathbf{w}^1_{\{2\}}  + \mathbf{D}_{3,\{1,2\}} \ \mathbf{w}^2_{\{1,2\}},\label{eq:user 3 needs to recover} \\
\mathbf{B}_4:=\mathbf{D}_{4,\{1\}} \ \mathbf{w}^1_{\{1\}}  + \mathbf{D}_{4,\{2,3\}} \ \mathbf{w}^2_{\{2,3\}},\label{eq:user 4 needs to recover} \\
\mathbf{B}_5:=\mathbf{D}_{5,\{2\}} \ \mathbf{w}^1_{\{2\}}  + \mathbf{D}_{5,\{1,3\}} \ \mathbf{w}^2_{\{1,3\}},\label{eq:user 5 needs to recover} \\
\mathbf{B}_6: =\mathbf{D}_{6,\{1\}} \ \mathbf{w}^1_{\{1\}}  + \mathbf{D}_{6,\{1,2\}} \ \mathbf{w}^2_{\{1,2\}}.\label{eq:user 6 needs to recover}
\end{align}

If we treat each sum in~\eqref{eq:user 1 needs to recover}-\eqref{eq:user 6 needs to recover} as a block and use the MAN strategy to delivery these blocks, 
we would transmit $B_1+B_2$, $B_3+B_4$, $B_5+B_6$ for a total of $\frac{\Fsf}{4}$ symbols.
Hence, the scheme achieves the same normalized load as   the   proposed scheme in~\eqref{eq:secondtolastcornerG} with $\mu_1=\frac{1}{2}$;  
in other words, a portion of the memory of size $\mu-\mu_1= \frac{47}{72}-\frac{1}{2}=\frac{11}{72}$ would be wasted. 
We next propose two novel schemes to let each user recover its desired sum in~\eqref{eq:user 1 needs to recover}-\eqref{eq:user 6 needs to recover}  while leveraging the whole memory.

{\it The solution that achieves $\rho_1$ in~\eqref{eq:r1}.} 
Focus on the demanded sum of user $1$ in~\eqref{eq:user 1 needs to recover}.
 The key idea is to   let user $1$ recover $\mathbf{D}_{1,\{2\}}  \mathbf{w}^1_{\{2\}} $  and  $\mathbf{D}_{1,\{2,3\}}   \mathbf{w}^2_{\{2,3\}}$  separately. In particular,
\begin{itemize}

\item
For the first term in $\mathbf{B}_1$ in~\eqref{eq:user 1 needs to recover}, since the dimension of $\mathbf{D}_{1,\{2\}}  $ is  $\frac{\Fsf}{12}   \times \frac{\Fsf}{24} $ and the sub-demand matrix $ \mathbf{D}_{1,\{2\}}  $ is known by  each  user, 
we let user $1$ directly recover $\mathbf{w}^1_{\{2\}}$, which contains  $\frac{\Fsf}{24} $ symbols, and then 
 compute  $\mathbf{D}_{1,\{2\}}  \mathbf{w}^1_{\{2\}} $.
Similarly, we let users $3,5$ recover $\mathbf{w}^1_{\{2\}}$, and  users $2,4,6$ recover $\mathbf{w}^1_{\{1\}}$. Thus in the delivery phase, the server transmits
\begin{align}
\mathbf{w}^1_{\{1\}}+\mathbf{w}^1_{\{2\}}, 
\end{align}
for a total of $\frac{\Fsf}{24} $ symbols, where users $1,3,5$ know $\mathbf{w}^1_{\{1\}}$ and users $2,4,6$ know $\mathbf{w}^1_{\{2\}}$.


\item 
For the second term in  $\mathbf{B}_1$ in~\eqref{eq:user 1 needs to recover}, since the dimension of $\mathbf{D}_{1,\{2,3\}}$ is  $ \frac{\Fsf}{12}   \times \frac{11\Fsf}{36} $ and the sub-demand matrix  $\mathbf{D}_{1,\{2,3\}}$ is known by  each   user, user $1$ needs to recover all symbols in  $\mathbf{D}_{1,\{2,3\}}   \mathbf{w}^2_{\{2,3\}}$. We denote $\mathbf{C}^{2}_{1,\{2,3,5,6\}} := \mathbf{D}_{1,\{2,3\}}   \mathbf{w}^2_{\{2,3\}}$  since it is known by users $2,3,5,6$. Hence, in order to let each user recover the first term in its desired sum,  the server transmits 
\begin{subequations}
\begin{align}
& \mathbf{C}^{2}_{1,\{2,3,5,6\}}+ \mathbf{C}^{2}_{2,\{1,3,4,6\}}  + \mathbf{C}^{2}_{3,\{1,2,4,5\}}, \\
& \mathbf{C}^{2}_{4,\{2,3,5,6\}}+ \mathbf{C}^{2}_{5,\{1,3,4,6\}}  + \mathbf{C}^{2}_{6,\{1,2,4,5\}},  
\end{align}
\end{subequations}
for a total of  $\frac{\Fsf}{6}$ symbols. 

\end{itemize}
Hence, in the delivery phase the server transmits $\frac{\Fsf}{24}  + \frac{\Fsf}{6}  =\frac{5\Fsf}{24}$ symbols, and the normalized load is $\rho_1=\frac{5}{24}$, which coincides with~\eqref{eq:r1}.

{\it The solution that achieves $\rho_2$ in~\eqref{eq:r2}.} 
 The idea is to partition each user's demand into two parts after having removed its cached content, where the partition is the result of a clever invertible linear transformation; we then have two steps, one for each of the two parts. 

We first focus on the demand of user $1$ in~\eqref{eq:user 1 needs to recover}, i.e.,  
\begin{align}
\mathbf{B}_1=\mathbf{D}_{1,\{2\}}  \mathbf{w}^1_{\{2\}}  + \mathbf{D}_{1,\{2,3\}}   \mathbf{w}^2_{\{2,3\}}= 
\left[ \begin{array}{c:c}
\mathbf{D}_{1,\{2\}} & \mathbf{D}_{1,\{2,3\}}  
\end{array}
\right]
\ \left[ \begin{array}{c}
 \mathbf{w}^1_{\{2\}} \\ \hdashline
\mathbf{w}^2_{\{2,3\}} 
\end{array} \right] \label{eq:example demand sum before trans user 1}.
\end{align}
The main strategy here is to take a linear transformations of~\eqref{eq:example demand sum before trans user 1} as follows
\begin{align}
\mathbf{B}^{\prime}_1= \left(\mathbf{T}_1\right)_{\frac{\Fsf}{12}  \times \frac{\Fsf}{12} } \  \left[ \begin{array}{c:c}
 \left( \mathbf{D}_{1,\{2\}} \right)_{\frac{\Fsf}{12}  \times \frac{\Fsf}{24} } & \left( \mathbf{D}_{1,\{2,3\}} \right)_{\frac{\Fsf}{12}  \times \frac{11\Fsf}{36} }  
\end{array} \right]  
 \  \left[ \begin{array}{c}
 \left( \mathbf{w}^1_{\{2\}} \right)_{\frac{\Fsf}{24}  \times 1} \\ \hdashline
\left(\mathbf{w}^2_{\{2,3\}}\right)_{\frac{11\Fsf}{36}  \times 1} 
\end{array} \right]  , \label{eq:example transformed demand of user 1}
\end{align}
where $\mathbf{T}_1$ is full-rank and the bottom $\frac{\Fsf}{12}-\frac{\Fsf}{24} =\frac{\Fsf}{24}$ symbols in $\mathbf{B}^{\prime}_1$ are linear combinations of $\mathbf{w}^2_{\{2,3\}}$  only  (i.e., these linear combinations do not contain any term in $\mathbf{w}^1_{\{2\}} $). This is possible because $\mathbf{B}_1$ contains $\frac{\Fsf}{12}$ linear combinations of all symbols in $[\mathbf{w}^1_{\{2\}} ; \mathbf{w}^2_{\{2,3\}}]$, while $\mathbf{w}^1_{\{2\}} $ contains $\frac{\Fsf}{24}$ symbols.
  Hence, we can re-express $\mathbf{B}^{\prime}_1$ as 
 \begin{align}
 \mathbf{B}^{\prime}_1 =\left[ \begin{array}{c}
 \left(  \mathbf{B}^{\prime}_{1,\{2,6\}} \right)_{\frac{\Fsf}{24}  \times 1} \\ \hdashline
\left(\mathbf{B}^{\prime}_{1,\{2,3,5,6\}}\right)_{\frac{\Fsf}{24}  \times 1} 
\end{array} \right], \label{eq:ex  transformed user 1}
 \end{align}
where  $ \mathbf{B}^{\prime}_{1,\{2,6\}}$ contains $\frac{\Fsf}{24}$ linear combinations of the symbols in $ \mathbf{w}^1_{\{2\}} $ and  $\mathbf{w}^2_{\{2,3\}} $,
which are both known by users $2$ and $6$,
while 
$ \mathbf{B}^{\prime}_{1,\{2,3,5,6\}}$ contains $\frac{\Fsf}{24}$ linear combinations of the symbols in $\mathbf{w}^2_{\{2,3\}}$ which are known by users in $\{2,3,5,6\}$. It will be clarified later that  the server transmits 
$ \mathbf{B}^{\prime}_{1,\{2,6\}}$  with coded caching gain equal to $g=2$ (i.e., the multicast message satisfies two users simultaneously), and 
$ \mathbf{B}^{\prime}_{1,\{2,3,5,6\}}$ with coded caching gain equal to $g+1=3$. 

Following the same line or reasoning, we can express the demands of the other users as 
\begin{align}
\mathbf{B}^{\prime}_2= [\mathbf{B}^{\prime}_{2,\{1,3\}};\mathbf{B}^{\prime}_{2,\{1,3,4,6\}} ];\label{eq:ex  transformed user 2} \\
\mathbf{B}^{\prime}_3= [\mathbf{B}^{\prime}_{3,\{2,4\}};\mathbf{B}^{\prime}_{3,\{1,2,4,5\}} ];\label{eq:ex  transformed user 3} \\
\mathbf{B}^{\prime}_4= [\mathbf{B}^{\prime}_{4,\{3,5\}};\mathbf{B}^{\prime}_{4,\{2,3,5,6\}} ];\label{eq:ex  transformed user 4} \\
\mathbf{B}^{\prime}_5= [\mathbf{B}^{\prime}_{5,\{4,6\}};\mathbf{B}^{\prime}_{5,\{1,3,4,6\}} ];\label{eq:ex  transformed user 5} \\
\mathbf{B}^{\prime}_6= [\mathbf{B}^{\prime}_{6,\{1,5\}};\mathbf{B}^{\prime}_{6,\{1,2,4,5\}} ].\label{eq:ex  transformed user 6}
\end{align} 
The transmission contains two steps. 
\begin{itemize}

\item
In the first step, we let each user $k\in [6]$ recover the first term of its demand $\mathbf{B}^{\prime}_k$.  In this step,  the server  transmits 
  \begin{subequations}
 \begin{align}
 &\mathbf{B}^{\prime}_{1,\{2,6\}} + \mathbf{B}^{\prime}_{2,\{1,3\}} ,\\
  &\mathbf{B}^{\prime}_{3,\{2,4\}} + \mathbf{B}^{\prime}_{4,\{3,5\}} ,\\
  &\mathbf{B}^{\prime}_{5,\{4,6\}}+\mathbf{B}^{\prime}_{6,\{1,5\}},
 \end{align}
  \end{subequations}
  which contains $\frac{ \Fsf}{8} $ symbols. 

\item
In the second step, we let each user $k\in [6]$ recover the second term of its demand $\mathbf{B}^{\prime}_k$. In this step, the server transmits 
  \begin{subequations}
 \begin{align}
 &\mathbf{B}^{\prime}_{1,\{2,3,5,6\}} + \mathbf{B}^{\prime}_{2,\{1,3,4,6\}} +  \mathbf{B}^{\prime}_{3,\{1,2,4,5\}},  \\
  &\mathbf{B}^{\prime}_{4,\{2,3,5,6\}} +\mathbf{B}^{\prime}_{5,\{1,3,4,6\}}  + \mathbf{B}^{\prime}_{6,\{1,2,4,5\}} , 
 \end{align}
  \end{subequations}
for a total of $\frac{\Fsf}{12} $ symbols. 
From the received multicast messages and its cache content, each user $k\in [\Ksf]$ can recover  $\mathbf{B}^{\prime}_k$, and then  compute  $\mathbf{B}_k$ from $\mathbf{T}^{-1}_k \mathbf{B}^{\prime}_k$.
\end{itemize}
 The normalized load is $\rho_2=\frac{ 1}{8}  + \frac{ 1}{12} =\frac{ 5}{24 } $, which conincides with~\eqref{eq:r2}. 
 
 In conclusion, the  normalized load of the proposed scheme is $\rho_{\text{ach}}=\min\{\rho_1,\rho_2 \}=\frac{ 5}{24 } $, while the baseline scheme in~\eqref{eq:first baseline load} achieves the normalized load equals $\frac{25}{72}$.
\hfill $\square$ 
\end{example}

\section{Numerical Evaluations}
\label{sec:numerical}
\begin{figure}
    \centering
    \begin{subfigure}[t]{0.5\textwidth}
        \centering
        \includegraphics[scale=0.6]{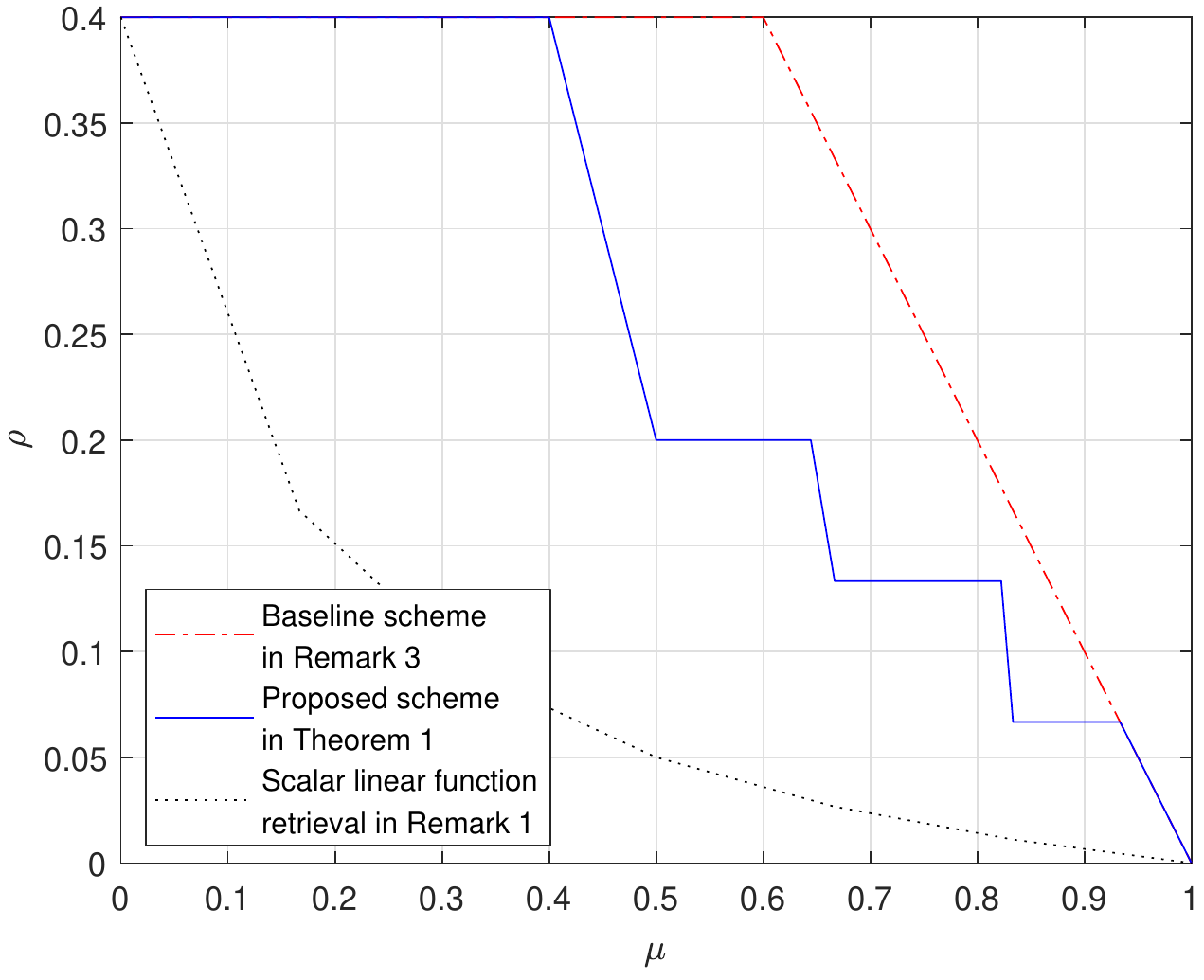}
        \caption{\small $\lambda=\frac{1}{15}$.}
        \label{fig:numerical 1a}
    \end{subfigure}%
    ~ 
    \begin{subfigure}[t]{0.5\textwidth}
        \centering
        \includegraphics[scale=0.6]{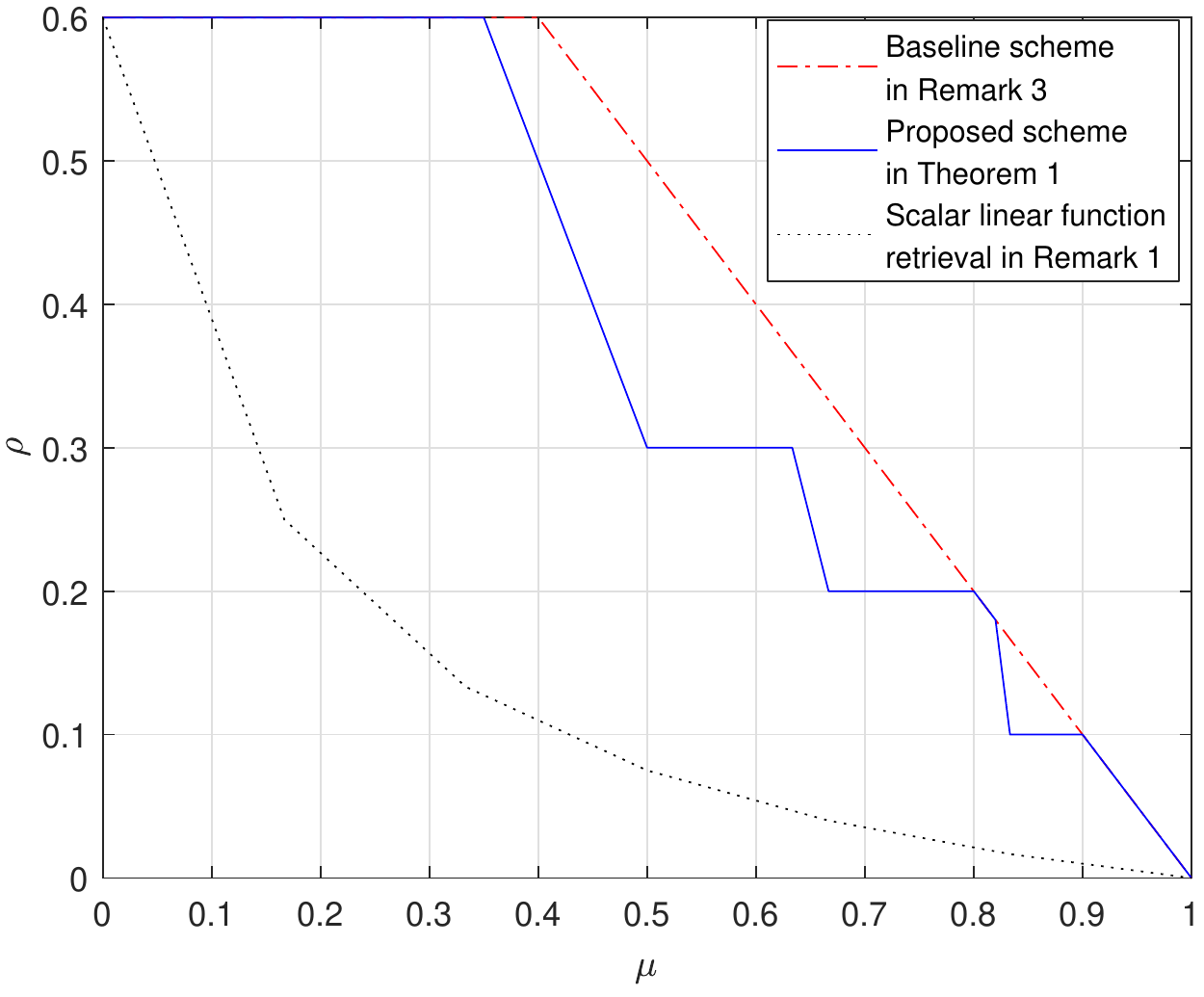}
        \caption{\small $\lambda=\frac{1}{10}$.}
        \label{fig:numerical 1b}
    \end{subfigure}
    \caption{\small  The   shared-link cache-aided general linear function retrieval problem $\Ksf=6$.  
    }
    \label{fig:num}
\end{figure} 

\begin{figure}
    \centering
    \begin{subfigure}[t]{0.5\textwidth}
        \centering
        \includegraphics[scale=0.6]{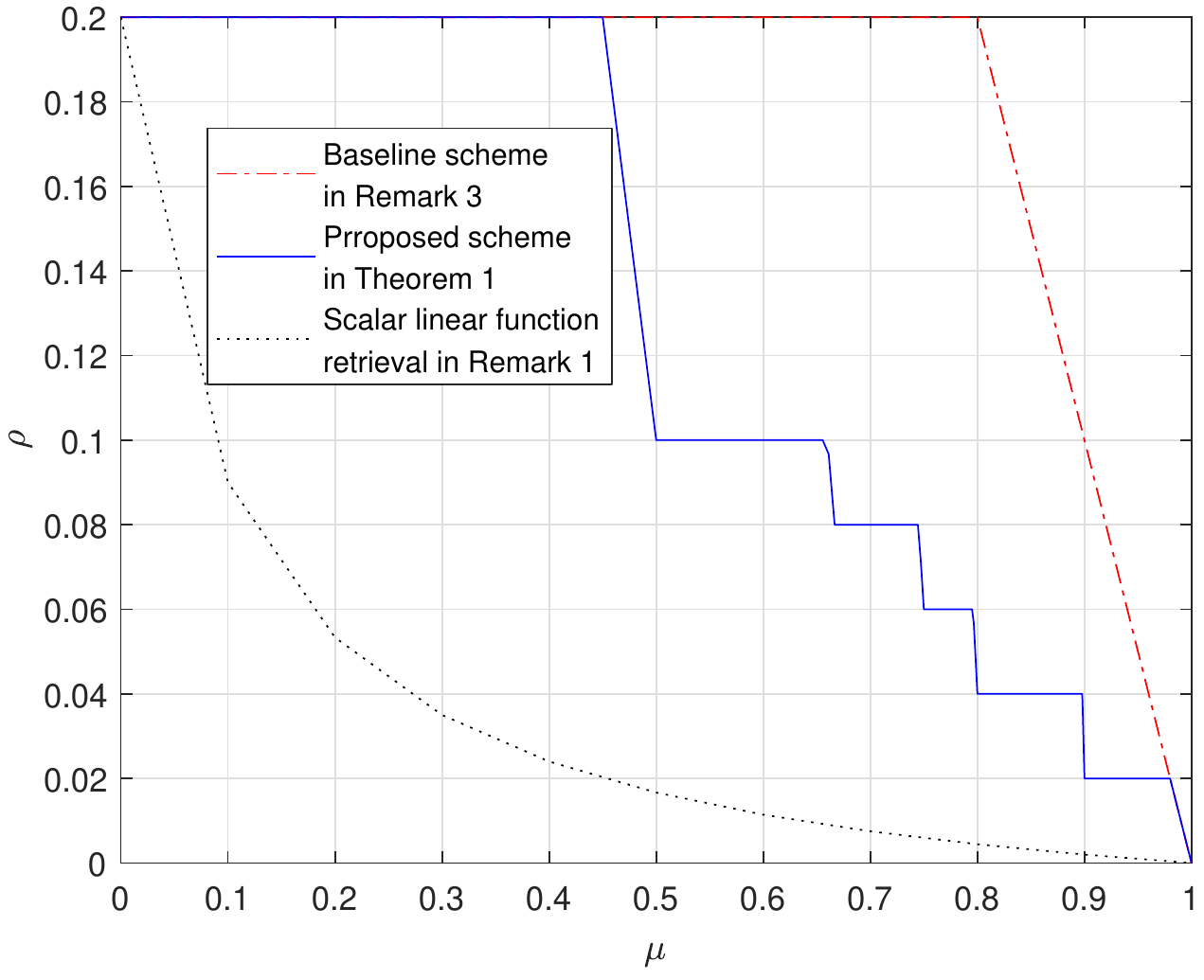}
        \caption{\small $\lambda=\frac{1}{50}$.}
        \label{fig:numerical 2a}
    \end{subfigure}%
    ~ 
    \begin{subfigure}[t]{0.5\textwidth}
        \centering
        \includegraphics[scale=0.6]{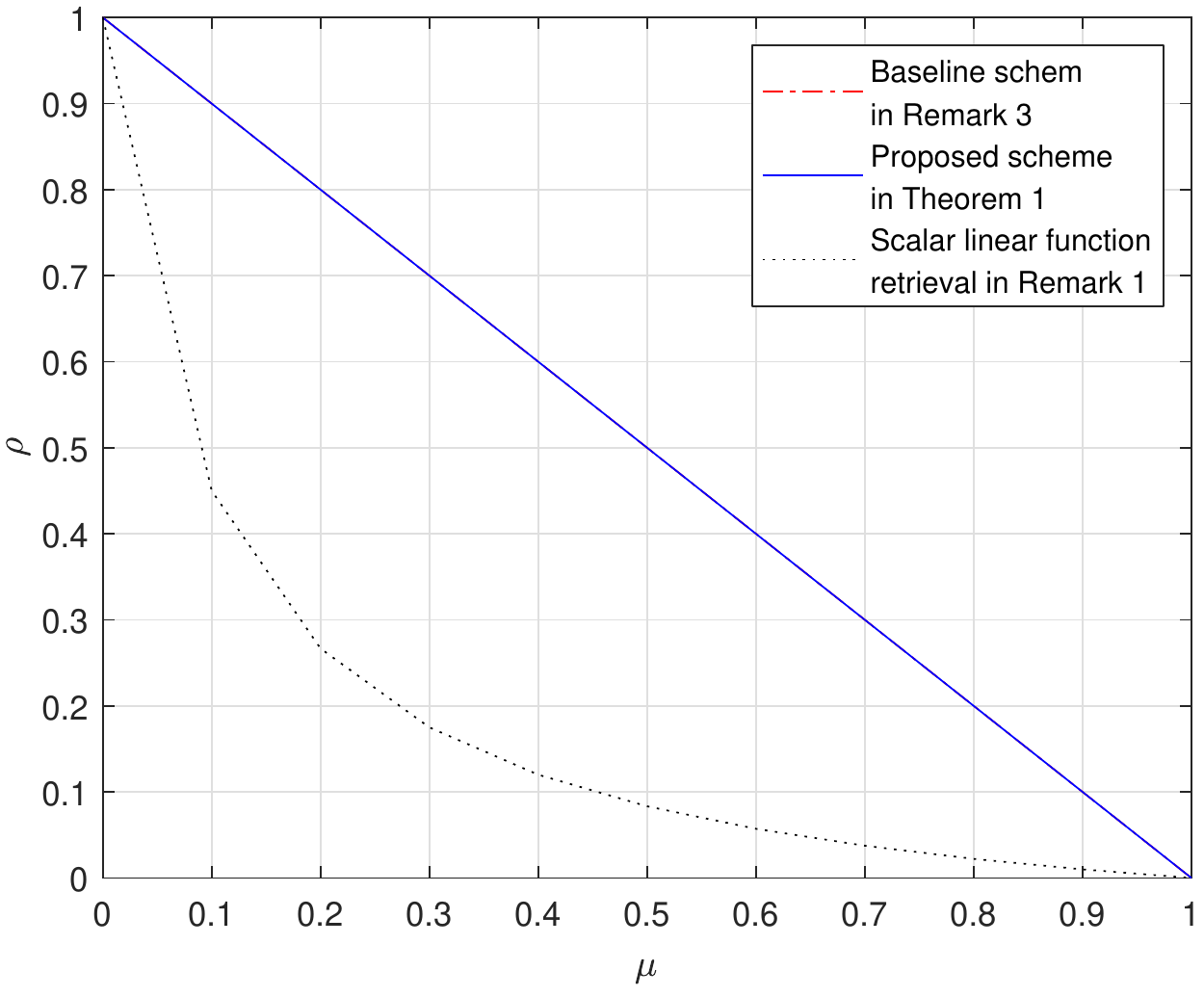}
        \caption{\small $\lambda=\frac{1}{10}$.}
        \label{fig:numerical 2b}
    \end{subfigure}
    \caption{\small  The   shared-link cache-aided general linear function retrieval problem $\Ksf=10$.  
    }
    \label{fig:num2}
\end{figure} 

We provide here some numerical evaluations on the performance of the proposed scheme in~\eqref{eq:mainthm}.
In Fig.~\ref{fig:numerical 1a} we consider the case $(\Ksf,\lambda)=(6, 1/15)$ and 
in Fig.~\ref{fig:numerical 1b} the case $(\Ksf,\lambda)=(6, 1/10)$. 
In  Fig.~\ref{fig:numerical 2a} we consider the case $(\Ksf,\lambda)=(10, 1/50)$ and 
in Fig.~\ref{fig:numerical 2b} the case $(\Ksf,\lambda)=(10, 1/10)$. 
From the figures, we observe that:
\begin{itemize}
\item 
In both settings our proposed scheme in Theorem~\ref{thm:main achievable scheme} outperforms the baseline scheme, as proved in Remark~\ref{rem:better than baseline}.
\item 
Fix $\Ksf$ and $\mu$. 
When $\lambda$ grows, the gap between the proposed scheme and the baseline scheme reduces.
When $\lambda= \frac{1}{\Ksf}$,  the proposed scheme  and the baseline scheme have the same load; this is because 
at the corner points of the proposed scheme in~\eqref{eq:secondtolastcornerG} we achieve the load $1-\mu$ which is the same as the baseline scheme. 
\item 
In addition, 
we also plot the cache-aided scalar linear function retrieval scheme in~\eqref{eq:YMA rho}, which only works for the case where the demand matrices are with the form in~\eqref{eq:qazxsw}. This comparison shows that, if the demand matrices are structured, we can design caching schemes that leverage the special structure of the demands to achieve a load that is  no larger  than the load for the worst-case demands.   Moreover, the more the structure the more the gain compared to in~\eqref{eq:mainthm}.
\end{itemize}

\section{Conclusions}
\label{sec:conclusion}
In this paper, we formulated the cache-aided general linear  function retrieval problem, where each user requests some linear combinations of all the symbols in the library.
The formulated problem generalizes 
the cache-aided scalar linear function retrieval problem. 
We proposed a novel scheme that strictly improves on an uncoded caching baseline scheme.
Further directions include  designing improved coded caching schemes for arbitrary users' demand ranges  (the setting considered here), as well as for given specific users' demand ranges.
In addition, the derivation of a converse bound is also part of on-going work.

  \appendices

\section{Proof of Theorem~\ref{thm:main achievable scheme}}
\label{app:achieve}

By a grouping strategy, we can achieve the normalized memory-load points in~\eqref{eq:secondtolastcornerG}. 
In the following, inspired by Example~\ref{ex:example alpha>0},   we  introduce a general interpolation method among the   points in~\eqref{eq:secondtolastcornerG}. 
 
We   let  $\text{Mod} (b,a)$ represent  the modulo operation on $b$ with  integer divisor $a$ and 
we let $\text{Mod}(b,a)\in \{1,\ldots,a \}$ (i.e., we let $ \text{Mod}(b,a)=a$ if $a$ divides $b$).

\subsection{$g\in [\Ksf-1]$ and $\frac{\alpha }{g} \geq \left\lceil \frac{\Ksf}{g}\right\rceil \lambda$}
\label{subsub:first case}
 We first consider the case where $g\in [\Ksf-1]$ and  $\frac{\alpha  }{g} \geq \left\lceil \frac{\Ksf}{g}\right\rceil \lambda $. 
Recall that   $\mu=\alpha \frac{g-1}{g} +(1-\alpha) \frac{g}{g+1}  > \frac{g-1}{g} $.
 In this case, 
 we directly use the caching scheme in~\eqref{eq:secondtolastcornerG} for the memory size $\frac{g-1}{g} $ with achieved normalized load  
 \begin{align}
 \min\left\{\left\lceil \frac{\Ksf}{g}\right\rceil \lambda,\frac{1}{g} \right\}  =\left\lceil \frac{\Ksf}{g}\right\rceil \lambda,
 \end{align}
which coincides with~\eqref{eq:first reg proposed ache}.

\subsection{$g\in [\Ksf-1]$ and $\frac{\alpha  }{g} \leq \left\lceil \frac{\Ksf}{g}\right\rceil \lambda$}
\label{subsub:second case}
We then focus on the case   where $g\in [\Ksf-1]$ and  $\frac{\alpha  }{g} \leq \left\lceil \frac{\Ksf}{g}\right\rceil \lambda $.

{\it  Placement Phase.}  
The placement is done by the memory-sharing between the proposed placements in~\eqref{eq:secondtolastcornerG} for $\Msf_1= \frac{g-1}{g} $ and $\Msf_2=\frac{g}{g+1} $.
We  divide    $\mathbf{w}$ into two parts, $\mathbf{w}=[\mathbf{w}^1;\mathbf{w}^2]$ where 
the dimension of   $\mathbf{w}^1$ is $\alpha \Fsf \times 1$ and the dimension  $\mathbf{w}^2$ is  $(1-\alpha)\Fsf \times 1 $. 

For the first part,  we further partition 
  $\mathbf{w}^1$   into $g$ non-overlapping and equal-length subfiles, $\mathbf{w}^1=[\mathbf{w}^1_{\Tc}: \Tc\subseteq [g] , |\Tc|=g-1 ]$, where 
the dimension of each subfile    $\mathbf{w}^1_{\Tc}$ is $\frac{ \alpha \Fsf}{g} \times 1$. 
  Each user $k\in [\Ksf]$ caches $\mathbf{w}^1_{\Tc}$ where   $\Tc\subseteq [g]$, $|\Tc| =g-1$, and $\text{Mod}(k,g)\in \Tc$.

  For the second part of each file, we further partition 
  $\mathbf{w}^2$   into $g+1$ non-overlapping and equal-length subfiles, $\mathbf{w}^2=[\mathbf{w}^2_{\Tc}: \Tc\subseteq [G+1] , |\Tc|=g]$, where 
the dimension of each subfile    $\mathbf{w}^2_{\Tc}$ is  $\frac{ (1-\alpha) \Fsf}{g+1} \times 1$. 
  Each user $k\in [\Ksf]$ caches $\mathbf{w}^2_{ \Tc}$ where   $\Tc\subseteq [g+1]$, $|\Tc| =G$, and $\text{Mod}(k,g+1)\in \Tc$.

  In total, each user caches 
  \begin{align}
  (g-1) \frac{ \alpha \Fsf }{g}  +     \frac{ g (1-\alpha) \Fsf}{g+1}   = \mu \Fsf 
  \end{align}
  symbols, satisfying the memory size constraint. 
  
 {\it  Delivery Phase.}  
 For each $\Tc_1 \in [g]$ where $|\Tc_1|=g-1 $, we define $\mathbf{D}_{k,\Tc_1}$ as the sub-matrix of $ \mathbf{D}_k$ which contains the columns corresponding to the symbols in $\mathbf{w}^1_{\Tc_1}$. In addition, for each $\Tc_2 \in [g+1]$ where $|\Tc_2|=g $, we define $\mathbf{D}_{k,\Tc_2}$ as the sub-matrix of $ \mathbf{D}_k$ which contains the columns corresponding to the symbols in $\mathbf{w}^2_{\Tc_2}$.
 
We can express the demand of user $k\in [\Ksf]$ as 
  \begin{subequations}
\begin{align}
 \mathbf{D}_k \ \mathbf{w} &= \sum_{\Tc_1 \in [g]: |\Tc_1|=g-1} \mathbf{D}_{k,\Tc_1} \mathbf{w}^1_{\Tc_1} + \sum_{\Tc_2 \in [g+1]: |\Tc_2|=g} \mathbf{D}_{k,\Tc_2} \mathbf{w}^2_{\Tc_2} \\
 &=  \sum_{\substack{\Tc_1 \in [g]: |\Tc_1|=g-1,  \\ \text{Mod}(k,g ) \in \Tc_1}} \mathbf{D}_{k,\Tc_1} \mathbf{w}^1_{\Tc_1} +\sum_{\substack{\Tc_2 \in [g+1]: |\Tc_2|=g, \\ \text{Mod}(k,g+1) \in \Tc_2}} \mathbf{D}_{k,\Tc_2} \mathbf{w}^2_{\Tc_2}  \nonumber \\
& +\mathbf{D}_{k, [g]\setminus \{\text{Mod}(k,g)\} } \  \mathbf{w}^1_{[g]\setminus  \{\text{Mod}(k,g )\}} +  \mathbf{D}_{k, [g+1]\setminus \{\text{Mod}(k,g+1 ) \}} \  \mathbf{w}^2_{[g+1]\setminus \{\text{Mod}(k,g+1 )\}}. \label{eq:first repress user k}
\end{align} 
  \end{subequations}
 It can be seen that user $k$ knows all the terms in~\eqref{eq:first repress user k} except 
\begin{align}
 \mathbf{B}_k:= \mathbf{D}_{k, [g]\setminus \{\text{Mod}(k,g)\} } \  \mathbf{w}^1_{[g]\setminus  \{\text{Mod}(k,g)\}} +  \mathbf{D}_{k, [g+1]\setminus \{\text{Mod}(k,g+1 ) \}} \  \mathbf{w}^2_{[g+1]\setminus \{\text{Mod}(k,g+1 )\}}  . \label{eq:need to recovered by user k}
\end{align} 
 Hence, in the delivery phase user $k$ should recover $\mathbf{B}_k$.
 We then propose two solutions for this objective.

{\it The solution that achieves $\rho_1$ in~\eqref{eq:r1}.} 
 We let user $k$ recover 
\begin{subequations}
\begin{align}
& \mathbf{B}_{k,1}:=\mathbf{D}_{k, [g]\setminus \{\text{Mod}(k,g)\} } \  \mathbf{w}^1_{[g]\setminus  \{\text{Mod}(k,g )\}}, \label{eq:Bk1} \\
& \mathbf{B}_{k,2}:=\mathbf{D}_{k, [g+1]\setminus \{\text{Mod}(k,g+1 ) \}} \  \mathbf{w}^2_{[g+1]\setminus \{\text{Mod}(k,g+1 )\}}.\label{eq:Bk2}
\end{align} 
\end{subequations} 
 
 For the first term $\mathbf{B}_{k,1}$,   the dimension of $\mathbf{D}_{k, [g]\setminus \{\text{Mod}(k,g)\} }  $ is  $ \lambda \Fsf  \times \frac{ \alpha   \Fsf}{ g}$ and $\mathbf{D}_{k, [g]\setminus \{\text{Mod}(k,g)\} }  $  is known by each user. Recall that in this case we have $\frac{ \alpha   \Fsf}{ g} \leq \lambda\Fsf $. Hence, we let user $k$ directly recover $\mathbf{w}^1_{[g]\setminus  \{\text{Mod}(k,g )\}}$. 
Thus in the delivery phase, we let the server transmit
\begin{align}
\sum_{i\in [g]} \mathbf{w}^1_{[g]\setminus  \{i\} },\label{eq:general first solution first term}
\end{align} 
with $\frac{ \alpha   \Fsf}{ g}$ symbols. 
 It can be seen that each user $k\in [\Ksf]$ desires $\mathbf{w}^1_{[g]\setminus  \{\text{Mod}(k,g )\}}$ and caches all the other terms in~\eqref{eq:general first solution first term}, such that user $k$ can recover $\mathbf{w}^1_{[g]\setminus  \{\text{Mod}(k,g )\}}$.

 For the second term  $\mathbf{B}_{k,2}$, the dimension of $\mathbf{D}_{k, [g+1]\setminus \{\text{Mod}(k,g+1)\} }  $ is  $ \lambda \Fsf  \times \frac{ (1-\alpha)   \Fsf}{ g+1}$. Notice that   
 $\mathbf{B}_{k,2}$ only  contains linear combinations of the second parts of files in the library. For the second part of each file, the users in 
 $$
 \Gc^2_{i}=\{k\in [\Ksf]: \text{Mod}(k,g+1 )= i\}
 $$
cache the same content, where $i\in [g+1]$.
Thus we can use the proposed delivery scheme in~\eqref{eq:secondtolastcornerG}. More precisely, 
for each $i\in [g+1]$, we generate a virtual  user $v_i$ with the demand 
\begin{align}
\mathbf{D}^{\prime}_{i} \  \mathbf{w}^2_{[g+1]\setminus \{i\}} =
\left[ \begin{array}{c}
\mathbf{D}_{\Gc^2_{i} (1), [g+1]\setminus \{i\} } \\ \hdashline
\mathbf{D}_{\Gc^2_{i} (2), [g+1]\setminus \{i\} } \\ \hdashline
 \vdots\\ \hdashline
\mathbf{D}_{\Gc^2_{i} (| \Gc^2_{i}|), [g+1]\setminus \{i\} } 
\end{array} \right]  \ \mathbf{w}^2_{[g+1]\setminus \{i\}}.\label{eq:r1 virtual user demands}
\end{align}
Notice that the dimension of $\mathbf{D}^{\prime}_{i}$ is $| \Gc^2_{i}| \lambda \Fsf  \times \frac{ (1-\alpha)  \Fsf}{ g+1}$. So virtual  user $v_i$ only needs to recover at most  
$\min \left\{ \left\lceil \frac{\Ksf}{g+1}\right\rceil \lambda,  \frac{  1-\alpha }{ g+1} \right\} \Fsf$ symbols in~\eqref{eq:r1 virtual user demands}.  We denote the
set of these symbols by $\mathbf{P}^{\prime}_{i, [g+1]\setminus \{i\}}$, which is known by all the other virtual users.
We then let the server transmit 
\begin{align}
\sum_{i\in [g+1]} \mathbf{P}^{\prime}_{i, [g+1]\setminus \{i\}},
\end{align}
with $\min \left\{ \left\lceil \frac{\Ksf}{g+1}\right\rceil \lambda,  \frac{ 1-\alpha }{ g+1} \right\} \Fsf$ symbols, such that each virtual user can recover its demand. 

In total, the server transmits 
\begin{align}
\frac{ \alpha    \Fsf}{ g} +\min \left\{ \left\lceil \frac{\Ksf}{g+1}\right\rceil \lambda,  \frac{  1-\alpha }{ g+1} \right\} \Fsf  =\rho_1 \Fsf
\end{align}
symbols, which  coincides with~\eqref{eq:r1}.

{\it The solution that achieves $\rho_2$ in~\eqref{eq:r2}.} 
 Recall that the   demanded sum of user $k$ is 
   \begin{subequations}
\begin{align}
\mathbf{B}_k&= \mathbf{D}_{k, [g]\setminus \{\text{Mod}(k,g)\} } \  \mathbf{w}^1_{[g]\setminus  \{\text{Mod}(k,g )\}} +  \mathbf{D}_{k, [g+1]\setminus \{\text{Mod}(k,g+1 ) \}} \  \mathbf{w}^2_{[g+1]\setminus \{\text{Mod}(k,g+1 )\}} \\
&= \left[ \begin{array}{c:c}
 \mathbf{D}_{k, [g]\setminus \{\text{Mod}(k,g)\} } & \ \mathbf{D}_{k, [g+1]\setminus \{\text{Mod}(k,g+1 ) \}}  
\end{array} \right]  
 \  \left[ \begin{array}{c}
\mathbf{w}^1_{[g]\setminus  \{\text{Mod}(k,g )\}} \\ \hdashline
\mathbf{w}^2_{[g+1]\setminus \{\text{Mod}(k,g+1 )\}}
\end{array} \right].
\end{align}
  \end{subequations}
  We can take a linear transformations on $\mathbf{B}_k$ as follows
 \begin{align}
 \mathbf{B}^{\prime}_k=    \mathbf{T}_k   \ \left[ \begin{array}{c:c}
  \mathbf{D}_{k, [g]\setminus \{\text{Mod}(k,g)\} }  & \ \mathbf{D}_{k, [g+1]\setminus \{\text{Mod}(k,g+1 ) \}}  
\end{array} \right]  
 \  \left[ \begin{array}{c}
\mathbf{w}^1_{[g]\setminus  \{\text{Mod}(k,g )\}} \\ \hdashline
\mathbf{w}^2_{[g+1]\setminus \{\text{Mod}(k,g+1 )\}}
\end{array} \right]  , \label{eq:transformed demand of user k}
 \end{align}
  where $\mathbf{T}_k$ is full-rank  with dimension  $\lambda \Fsf \times \lambda \Fsf$,
  and the bottom  $\left[\lambda-  \frac{\alpha  }{g}\right]^+ \Fsf$ symbols in $\mathbf{B}^{\prime}_k$ are some linear combinations of $\mathbf{w}^2_{[g+1]\setminus \{\text{Mod}(k,g+1 )\}}$ (i.e., these linear combinations do not contain any term in $\mathbf{w}^1_{[g]\setminus  \{\text{Mod}(k,g )\}}$).
 This is possible because $\mathbf{B}_k$ contains $\lambda \Fsf$ linear combinations of all symbols in 
$[\mathbf{w}^1_{[g]\setminus  \{\text{Mod}(k,g )\}};\mathbf{w}^2_{[g+1]\setminus \{\text{Mod}(k,g+1 )\}}]$, while  
 $\mathbf{w}^1_{[g]\setminus  \{\text{Mod}(k,g )\}}$ contains $\frac{\alpha   \Fsf}{g}$ symbols.
  Hence, we can re-express $\mathbf{B}^{\prime}_k$ as 
 \begin{align}
 \mathbf{B}^{\prime}_k =\left[ \begin{array}{c}
  \left(  \mathbf{B}^{\prime}_{k,1} \right)_{\min\left\{\frac{\alpha   \Fsf}{g},\lambda \Fsf \right\} \times 1} \\ \hdashline
\left(\mathbf{B}^{\prime}_{k,2}\right)_{\left[\lambda -  \frac{\alpha }{g}\right]^+ \Fsf \times 1} 
\end{array} \right]. \label{eq:transformed user k}
 \end{align}
 
 The delivery phase is divided into two steps. 
In the first step, we first let each user $k\in [\Ksf]$ recover $ B^{\prime}_{k,1}$. 
 Notice that $ \mathbf{B}^{\prime}_{k,1}$ is the set of some linear combinations of the symbols in  $
\mathbf{w}^1_{[g]\setminus  \{\text{Mod}(k,g )\}} 
\mathbf{w}^2_{[g+1]\setminus \{\text{Mod}(k,g+1 )\}} 
 $. $\mathbf{w}^1_{[g]\setminus  \{\text{Mod}(k,g )\}} $ is known by any user $j_1 \in [\Ksf]$ where $\text{Mod}(j_1,g ) \neq k$;
 $\mathbf{w}^2_{[g+1]\setminus \{\text{Mod}(k,g+1 )\}} $ is known by any user $j_2\in [\Ksf]$ where $\text{Mod}(j_2,g+1 ) \neq k$. 
Assume that $k= a_k g+ \text{Mod}(k,g)$, where $a_k= \left\lceil \frac{k}{g}\right\rceil -1 $ and $\text{Mod}(k,g ) \in [g]$. 
 In Appendix~\ref{sec:division lemma}, we prove the following lemma.
 \begin{lem}
 \label{lemma:division lemma}
 Each user $k_1=a_k g+ j$ where $j \in [g] \setminus \{\text{Mod}(k,g )\} $ and $k_1\in [\Ksf]$,  caches both  $\mathbf{w}^1_{[g]\setminus  \{\text{Mod}(k,g )\}} $ and  $\mathbf{w}^2_{[g+1]\setminus \{\text{Mod}(k,g+1 )\}} $.
  \hfill $\square$ 
 \end{lem}

For each $i\in  \left[\left\lceil \frac{\Ksf}{g }\right\rceil \right]$, we let the server transmit 
\begin{align}
\sum_{j\in [g]: (i-1)g+j \leq \Ksf}  \mathbf{B}^{\prime}_{(i-1)g+j ,1}. \label{eq:r2 first round}
\end{align}
From Lemma~\ref{lemma:division lemma}, each user $(i-1)g+j $ knows all except $\mathbf{B}^{\prime}_{(i-1)g+j ,1}$ such that it can recover  $\mathbf{B}^{\prime}_{(i-1)g+j ,1}$. In this step, the server   transmits 
$
\left\lceil \frac{\Ksf}{g }\right\rceil \min\left\{\frac{\alpha  }{g},\lambda \right\}  \Fsf
$
symbols.

In the second step,
we then let each user $k\in [\Ksf]$ recover $ \mathbf{B}^{\prime}_{k,2}$, which contains linear combinations of $\mathbf{w}^2_{[g+1]\setminus \{\text{Mod}(k,g+1 )\}} $. 
  We can use the same delivery scheme as we used to delivery the second term  in the first solution (i.e., $\mathbf{B}_{k,2}$ in~\eqref{eq:Bk2} which contains $\lambda\Fsf$ linear combinations of $\mathbf{w}^2_{[g+1]\setminus \{\text{Mod}(k,g+1 )\}} $). 
Here we do not repeat the scheme. Notice that 
$ \mathbf{B}^{\prime}_{k,2}$ contains $ \left[\lambda-  \frac{\alpha  }{g}\right]^+ \Fsf$ linear combinations of $\mathbf{w}^2_{[g+1]\setminus \{\text{Mod}(k,g+1 )\}} $. Hence, in this step the server   transmits 
$ \min \left\{ \left\lceil \frac{\Ksf}{g+1}\right\rceil \left[\lambda-  \frac{\alpha  }{g}\right]^+,  \frac{  1-\alpha }{ g+1} \right\} \Fsf$ symbols.

After recovering  $ \mathbf{B}^{\prime}_k$,  each user $k\in [\Ksf]$ reconstructs 
$$
  \mathbf{B}_k =  \mathbf{T}^{-1}_k  \mathbf{B}^{\prime}_k,
 $$
and then recovers its demand.

In total, the achieved normalized load is 
\begin{align*}
\left\lceil \frac{\Ksf}{g }\right\rceil  \min\left\{\frac{\alpha  }{g},\lambda \right\}  +  \min \left\{ \left\lceil \frac{\Ksf}{g+1}\right\rceil \left[\lambda-  \frac{\alpha  }{g}\right]^+,  \frac{  1-\alpha }{ g+1} \right\}=\rho_2,
\end{align*} 
 coinciding with~\eqref{eq:r2}.

\subsection{Proof of~\eqref{eq:sec reg proposed ache dt3}}
\label{sub:third case proposed scheme} 
 Finally, we focus on the case where 
  $\mu =\alpha \frac{\Ksf-1}{\Ksf} +(1-\alpha) $ where $\alpha \in (0,1)$. 
  In this case, the proposed scheme is a direct extension from the proposed scheme  in~\eqref{eq:secondtolastcorner K-1}.
  More precisely, 
\begin{itemize}
\item  we directly use the caching scheme in~\eqref{eq:secondtolastcorner K-1} for the memory size $\frac{\Ksf-1}{\Ksf} $ with the achieved normalized load equal to $\lambda$.
\item  In this case,   the number of symbols which are not cached by user is $\frac{\alpha  \Fsf}{\Ksf}$. Hence, we can let each user directly recover  the uncached symbols with the achieved normalized load equal to $\frac{\alpha }{\Ksf}$.
\end{itemize}  
This concludes the proof.

\section{Proof of Lemma~\ref{lemma:division lemma}}
\label{sec:division lemma}
Recall that  $k= a_k g+ \text{Mod}(k,g)$, where $a_k= \left\lceil \frac{k}{g}\right\rceil -1 $ and $\text{Mod}(k,g ) \in [g]$. We focus on one user 
 $k_1=a_k g+ j$ where $j \in [g] \setminus \{\text{Mod}(k,g )\} $.
 Since $j=\text{Mod}(k_1,g ) \neq \text{Mod}(k,g )$, it can be easily seen that $\mathbf{w}^1_{[g]\setminus  \{\text{Mod}(k,g )\}} $ is cached by user $j$.
In the rest of this proof, we show that user $j$ also caches  $\mathbf{w}^2_{[g+1]\setminus \{\text{Mod}(k,g+1 )\}} $; or equivalently, $\text{Mod}(k_1,g+1 ) \neq \text{Mod}(k,g+1 )$.

We prove it by contradiction. Assume that $\text{Mod}(k_1,g+1 ) = \text{Mod}(k,g+1 )=j^{\prime}$. Hence, we can re-express $k$ as $k=a^{\prime}_k (g+1) + j^{\prime}$ and re-express $k_1$ as $k_1= a^{\prime}_{k_1} (g+1)+j^{\prime}$, where $a^{\prime}_k=\left\lceil \frac{k}{g+1}\right\rceil -1 $
and $a^{\prime}_{k_1} =a^{\prime}_k=\left\lceil \frac{k_1}{g+1}\right\rceil -1 $.

Since $k=a_k g+ \text{Mod}(k,g) =a^{\prime}_k (g+1) + j^{\prime}$, we have 
\begin{align}
a_k g=a^{\prime}_k (g+1) + j^{\prime}-\text{Mod}(k,g).\label{eq:from k}
\end{align}
In addition, we have 
\begin{align}
k_1=a_k g+ j=a^{\prime}_{k_1} (g+1)+j^{\prime} \label{eq:from k_1}
\end{align}
By taking~\eqref{eq:from k} into~\eqref{eq:from k_1}, we have 
\begin{align}
& a_k g+ j=a^{\prime}_{k_1} (g+1)+j^{\prime}\\
&  \stackrel{\eqref{eq:from k}}{\Longleftrightarrow }   a^{\prime}_k (g+1) + j^{\prime}-\text{Mod}(k,g) + j=a^{\prime}_{k_1} (g+1)+j^{\prime} \\
&\Longleftrightarrow  ( a^{\prime}_k -a^{\prime}_{k_1})(g+1)= \text{Mod}(k,g) - j. \label{eq:contradicts}
\end{align}
Since $\text{Mod}(k,g) \in [g]$ and $j \in [g]$, it can be seen that~\eqref{eq:contradicts} holds if and only if $a^{\prime}_k -a^{\prime}_{k_1}=0$, which leads to $k=k_1$ and contradicts with  $\text{Mod}(k,g) \neq \text{Mod}(k_1,g)$. Hence, we proved that 
 $\text{Mod}(k_1,g+1 ) \neq  \text{Mod}(k,g+1 ) $ and proved Lemma~\ref{lemma:division lemma}.

\bibliographystyle{IEEEtran}
\bibliography{IEEEabrv,IEEEexample}
\end{document}